\documentclass[draft]{elsart}
\usepackage{psfig}
\newcommand\BE{\begin{equation}}
\newcommand\EE{\end{equation}}

\begin{document}
\begin{frontmatter}
\title{Fractal tracer distributions in turbulent field theories}

\author{Jonas Lundbek Hansen} and
\author{Tomas Bohr} 

\address{
Center for Chaos and Turbulence Studies \\
The Niels Bohr Institute, Blegdamsvej 17, 2100 Copenhagen
{\O}, 
Denmark \\
}
\begin{abstract}
We study the motion of passive tracers in a two-dimensional turbulent
velocity field generated by the Kuramoto-Sivashinsky equation.
By varying the direction of the velocity-vector with respect to the
field-gradient
we can continuously vary the two
Lyapunov exponents for the particle motion
and thereby find a regime in which the particle distribution is a
strange attractor.
We compare the Lyapunov dimension to the information dimension of
actual particle distributions and show that there is good agreement
with the Kaplan-Yorke conjecture.
Similar phenomena have been observed experimentally.
\end{abstract}

\end{frontmatter}

\section{Introduction}
\begin{figure}[t]
\centerline{
\mbox{
\psfig{figure=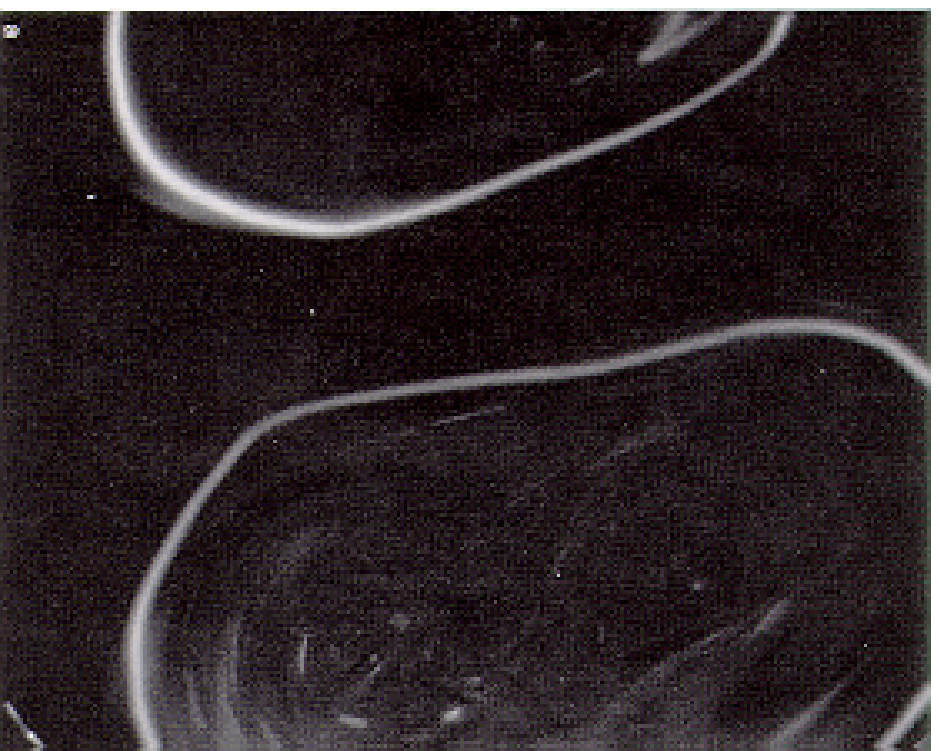,height=8cm,width=6.5cm}
\psfig{figure=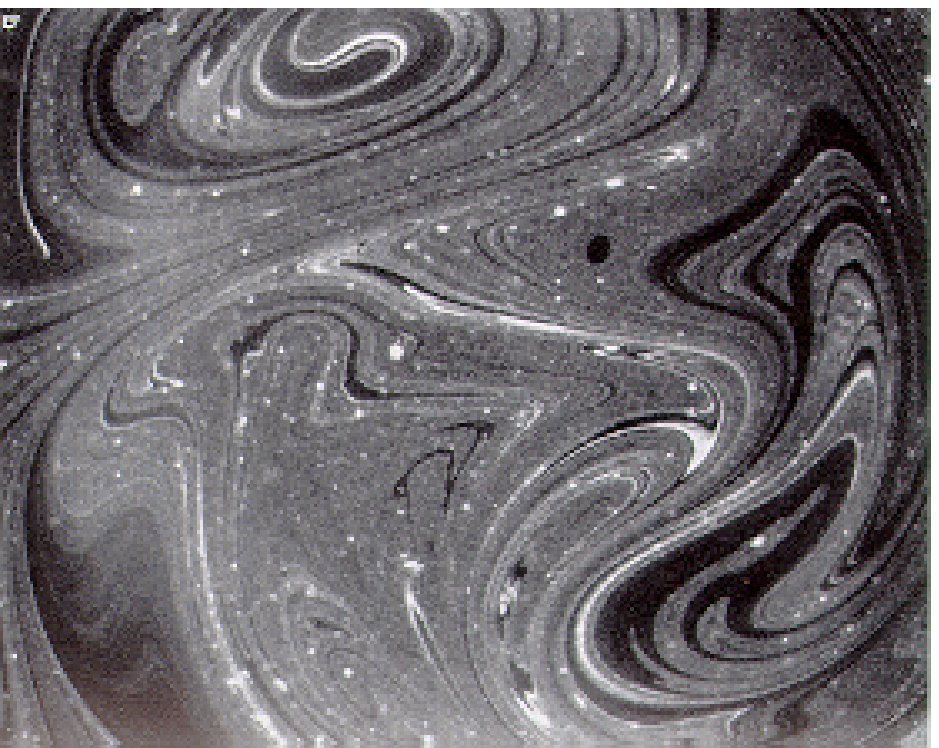,height=8cm,width=6.5cm}
}}
\caption[
Pictures of tracer distributions from experiments.
Left:  laminar case. Right:  fractal case.
From \protect{\sl Sommerer} [2]. 
{
Pictures of tracer distributions from experiments.
Left:  laminar case. Right:  fractal case.
From {\sl Sommerer} [2]. 
}
\label{fi:sommerer}
\end{figure}

\setlength{\baselineskip}{0.8cm}
The velocity field of particles confined to the surface of a 
 fluid does not have to be incompressible
 even though the fluid itself is incompressible. 
It has correspondingly been observed experimentally
\cite{sommererscience,sommerer}, that tracer particles moving only on the
 two-dimensional surface of a three-dimensional fluid can lie on a
 strange attractor as shown in FIG.~\ref{fi:sommerer}.
 In these experiments some $50$ million floating 
particles were advected on the
surface of the fluid, which was heavily stirred from time to time,
with the stirrings done so seldom, that the fluid had time to come to rest.
Pictures of the  particle distributions were then examined
and were found to
have a well-defined fractal dimension 
between 1 and 2 and hence to be a strange attractor,
which changed shape for each stirring.
This has been explained theoretically in terms of 
``random maps" \cite{LOS1,LOS2}, 
which seems appropriate for this pulsed flow. On the other hand, 
this phenomenon is rather general and should  also occur in systems  
with no separation between a stirring and relaxing phase. To 
investigate this possibility we have looked at the advection of passive 
scalars in a simple turbulent field theory, the Kuramoto-Sivashinsky (KS) 
equation, which has been derived e.g. in the context of 
chemical turbulence 
\cite{kuramoto} and the flow of a falling fluid film \cite{sivashinsky}.

\section{Advection by the 1D Kuramoto-Sivashinsky Equation}

The Kuramoto-Sivashinsky equation in 1+1 dimensions is given as 
\BE
 h_t  =
- h_{xxxx}  - h_{xx}  +(h_x)^2
\label{ks}
\EE
where $h$ is a scalar field and $f_x \equiv \frac{\partial f}{\partial x}$.
The boundary conditions are periodic on $x \in [0,L]$.
The equation for the derivative $u = h_x$ yields
\BE
 u_t  =
- u_{xxxx}  - u_{xx}  + 2 uu_x.
\label{uks}
\EE
This equation has the same nonlinear term as the Navier-Stokes
equations.
It is thus natural to look at the field $u=h_x$ as a velocity field and 
study the motion of a passive particle   
with the equations of motion $\frac{dx}{dt}= a h_x(x,t)$, 
where $a$ is a real constant that gives the ratio of the time scales
of the particle motion and the $h$-field.
Note that the $h$-field itself is not a good choice for a velocity field, since
the equation (\ref{ks}) 
is invariant under the transformation $h \rightarrow h+const$.

Such a study was carried out  by Bohr and Pikovsky \cite{ksdif}.
They examined the case $a=1$ and
found the particles to diffuse anomalously, i.e.,
the average displacement in time goes approximately as 
$<x(t)^2>^{\frac{1}{2}} \propto t^{2/3}$, 
which is faster than usual diffusion.
They also found, for systems seeded with several
particles, that particles coalesce in time.
Therefore the Lyapunov exponent for the 
particles is negative 
 even though they move in a chaotic  velocity field.
Thus, the system exhibits Eulerian but not Lagrangian chaos.
\section{Advection by the 2D Kuramoto-Sivashinsky Equation}
It is trivial to generalize the Kuramoto-Sivashinsky equation (\ref{ks})
to two dimensions. The equation is given as
\BE
{\partial  h  \over  \partial  t}  =
-{\bf \nabla} ^4  h  -{\bf \nabla} ^2  h  +({\bf \nabla} h) 
\cdot ({\bf \nabla} h).
\EE
The $h$-field is now
a function $h=h(x,y,t)$, and
${\bf \nabla} = {\bf i}\frac{\partial}{\partial x} +
{\bf j}\frac{\partial}{\partial y}$.
For the higher order derivatives, we use the
identities $\nabla^2 = \nabla \cdot \nabla$ and
$\nabla ^4 = \nabla \cdot \nabla ( \nabla \cdot \nabla)$.
The reason for the simplicity in the generalization is that all terms
are scalar products and hence scalars. Again we assume periodic
boundary conditions on $(x,y) \in [0,L] \times [0,L]$.

\begin{figure}[t]
\centerline{
\hbox{
\psfig{figure=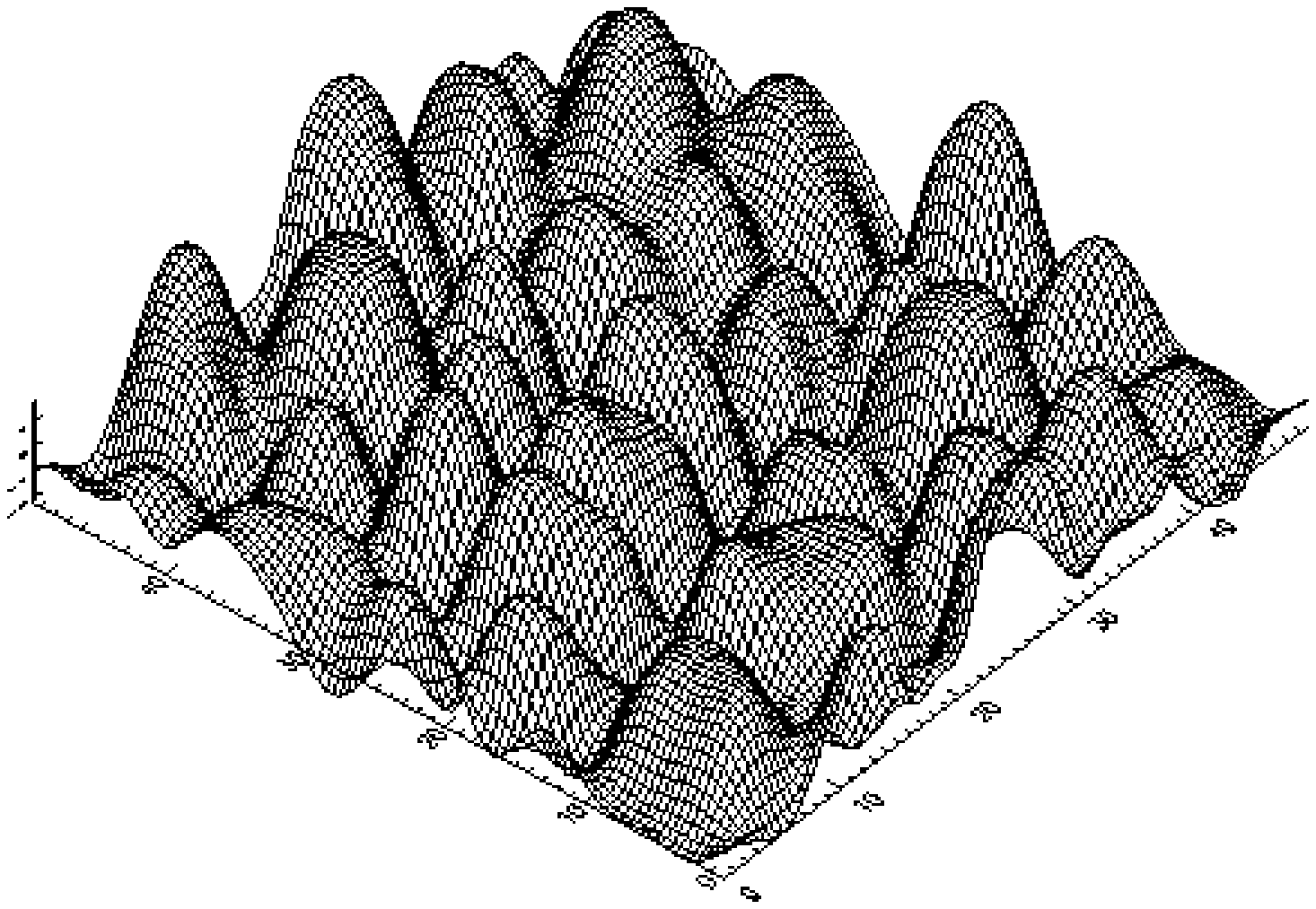,angle=270,height=6.5cm,width=6.5cm} 
\psfig{figure=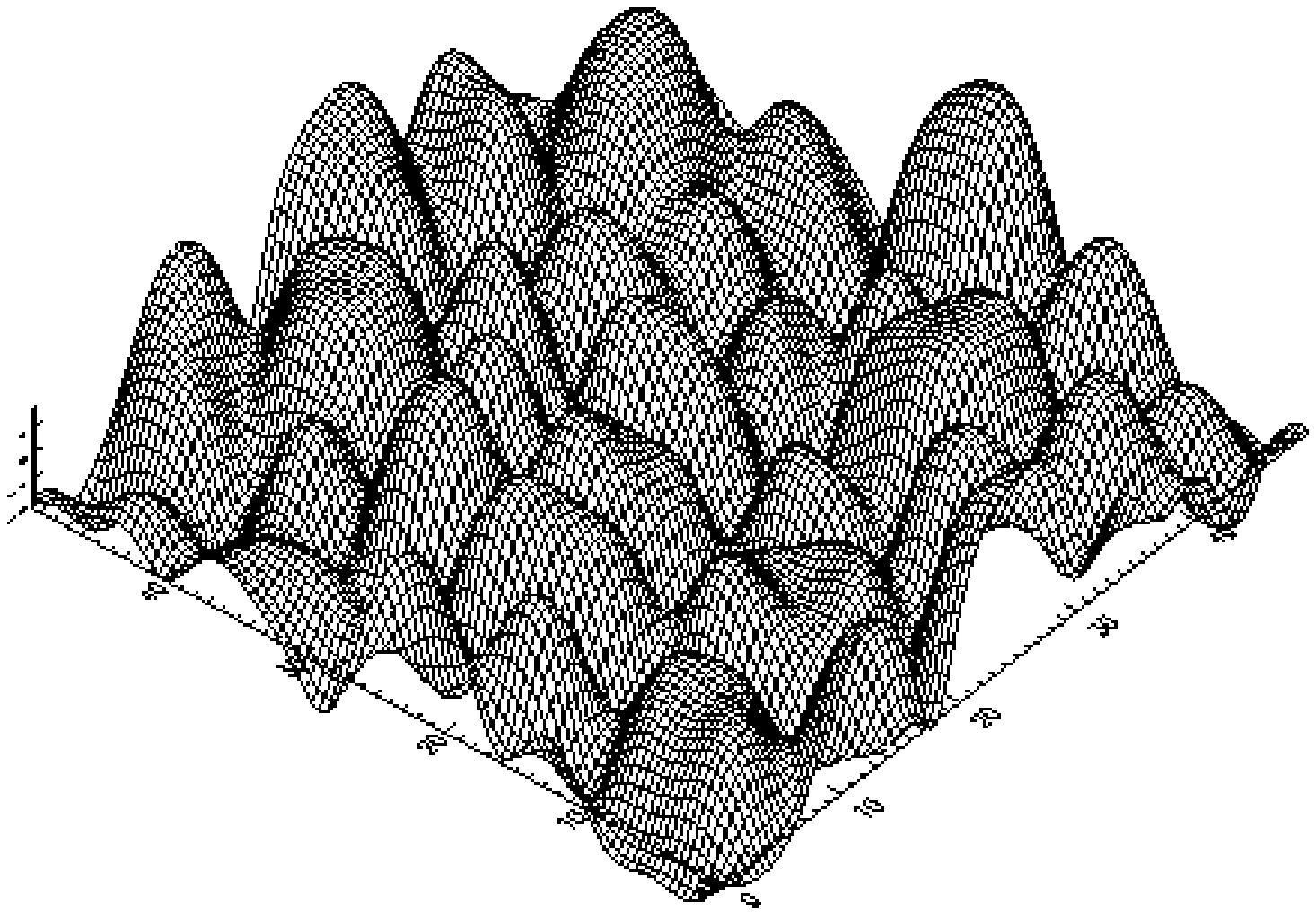,angle=270,height=6.5cm,width=6.5cm} 
}
}
\centerline{
\hbox{
\psfig{figure=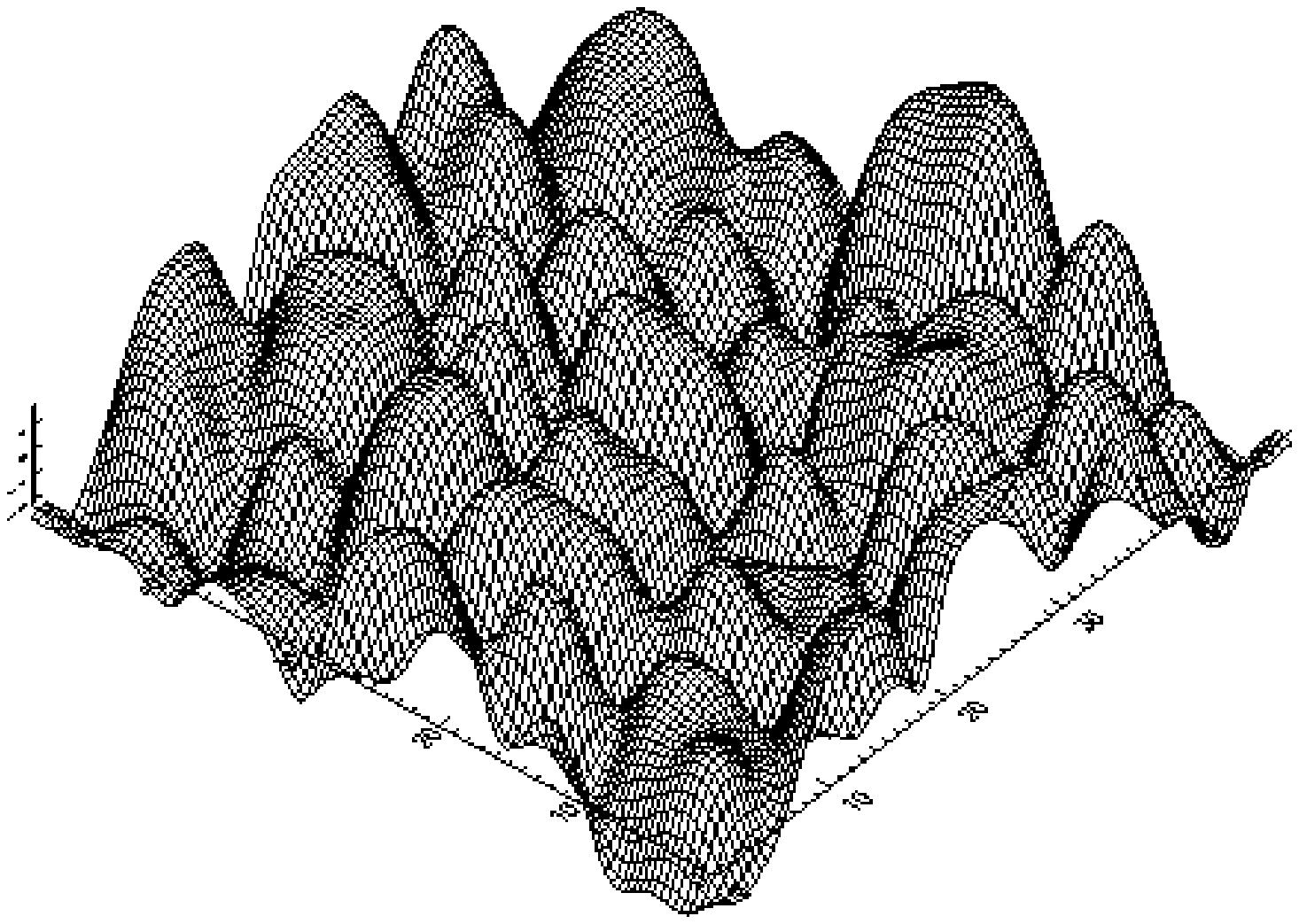,angle=270,height=6.5cm,width=6.5cm} 
\psfig{figure=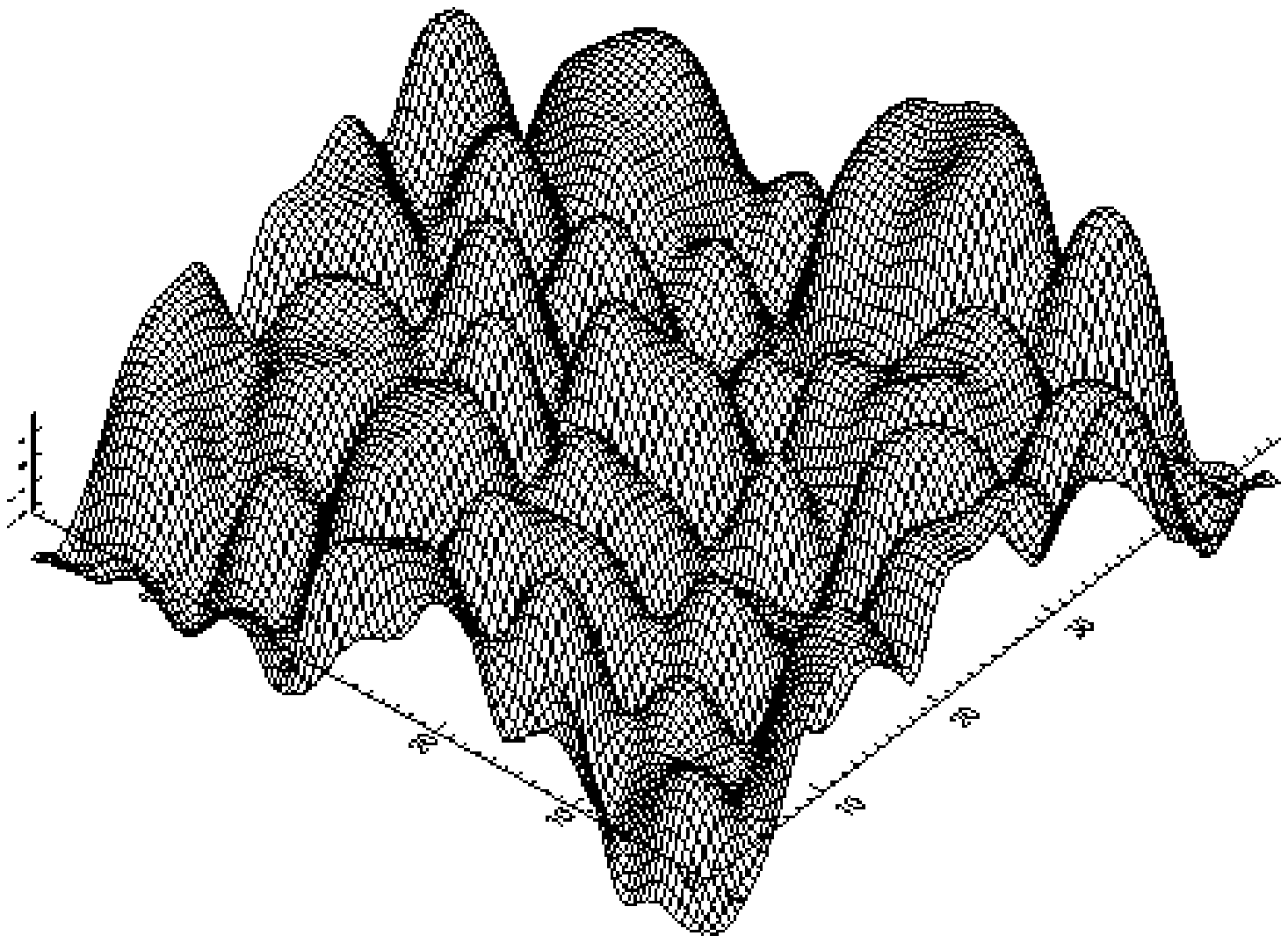,angle=270,height=6.5cm,width=6.5cm} 
}
}
\caption{
Solutions to the 2D Kuramoto-Sivashinsky equation at four consecutive times.
Time going with $\Delta t =1$ from upper left to lower right.
The system size is $50x50$.
}
\label{fig:ks2dsurf}
\end{figure}

With random initial conditions and periodic boundary conditions,
 the solutions look like mountain ranges:
mountains (maxima) are separated from each other by valleys (minima).
The valleys form a web where the mountains are isolated holes in the web.
In the course of the dynamics
mountains are created spontaneously or 
by splitting large mountains into (2,3 or 4) smaller ones -
or they can disappear by being ``squeezed'' by neighbouring mountains.
FIG.~\ref{fig:ks2dsurf} shows solutions to the 2D KS-equation 
with $L = 50$ at four consecutive time steps. 
\subsection{Choice of velocity field}
The most obvious choice of velocity field is, as in the 1D-case, 
a field proportional to the  gradients of  the $h$-field, i.e.,
\BE
{\bf v}_g = a {\bf \nabla}h 
\label{grad1}
\EE
where the real constant $a$, as in the 1D-case, is the ratio of 
the time scales of the particle motion and the $h$-field.
 The gradient always points in the direction of the local maxima. This
 causes particles with 
 different initial positions to coalesce at these maxima,
similar to what is seen in the 1D-case \cite{ksdif}. 
Thus, particles moving in the gradient velocity field have
 both Lyapunov exponents $\lambda_1,\lambda_2$ less than zero.
The original field $h$ can be regarded as a {\it velocity potential}.
This motion of particles is area contracting and hence dissipative.
 In terms of fluid dynamics, the velocity field ${\bf \nabla} h$ is
{\it compressible}.

We can generalize the equations of motion (\ref{grad1}) 
by a linear function of the gradient vectors.
Since we want to preserve the isotropy of the KS-equation,
the only possibility for the generalization is scaling and rotation.
The generalized velocity field is thus constructed as a 
multiplication of  a constant $a$,  
the rotation matrix ${\bf B}_{\theta}$
and the gradient vectors: 
\BE
{\bf v} = 
a 
{\bf B}_{\theta} 
\nabla h \quad\mbox{where}\quad           
{\bf B}_{\theta} 
=
\left( 
\begin{array}{cc} 
\cos{\theta} & \sin{\theta} \\
-\sin{\theta} & \cos{\theta} \\
\end{array}
\right)  
\EE 
 Of special interest is the vector rotated 
$90^{\circ}$ from the gradient:
\BE
{\bf v}_h
 =a 
{\bf B}_{90^{\circ}} 
\nabla h
 =
 \left(
 \begin{array}{c} 
  - h_y \\
  h_x \end{array}
  \right).
   \EE
This vector field viewed as a velocity field is {\sl incompressible}, i.e.
$\nabla \cdot {\bf v}_h = 0$, and the scalar field $h$ can be regarded as 
a {\sl stream function}.
 Equivalently, particle motion in this velocity field
 can be said to be a non-autonomous
 (time-dependent) Hamiltonian system with one degree of freedom,
 where $x$ and $y$ are conjugate variables and
 $h$ is the Hamiltonian.
 A Hamiltonian system is area preserving
 and hence the two Lyapunov exponents have the same numerical
 value with opposite signs,
 $(\lambda_1 = - \lambda_2)$. Thus, if a volume element is expanding in
one direction it should contract in the other to preserve the
volume \cite{ott}. 
Advected in this velocity field, the particles move around the
maxima of the field at a nearly constant height. The maxima can be regarded
 as vortices of the velocity field.

Throughout this paper,
we restrict ourselves to examining the behavior of particles with the most
natural ratio of the time scales, thus $a = 1$.
We then describe the vector field ${\bf v}$ by just one parameter, the angle
$\theta$.

 \subsection{Particle Trajectories}
\begin{figure}[t]
\centerline{
\hbox{
\psfig{figure=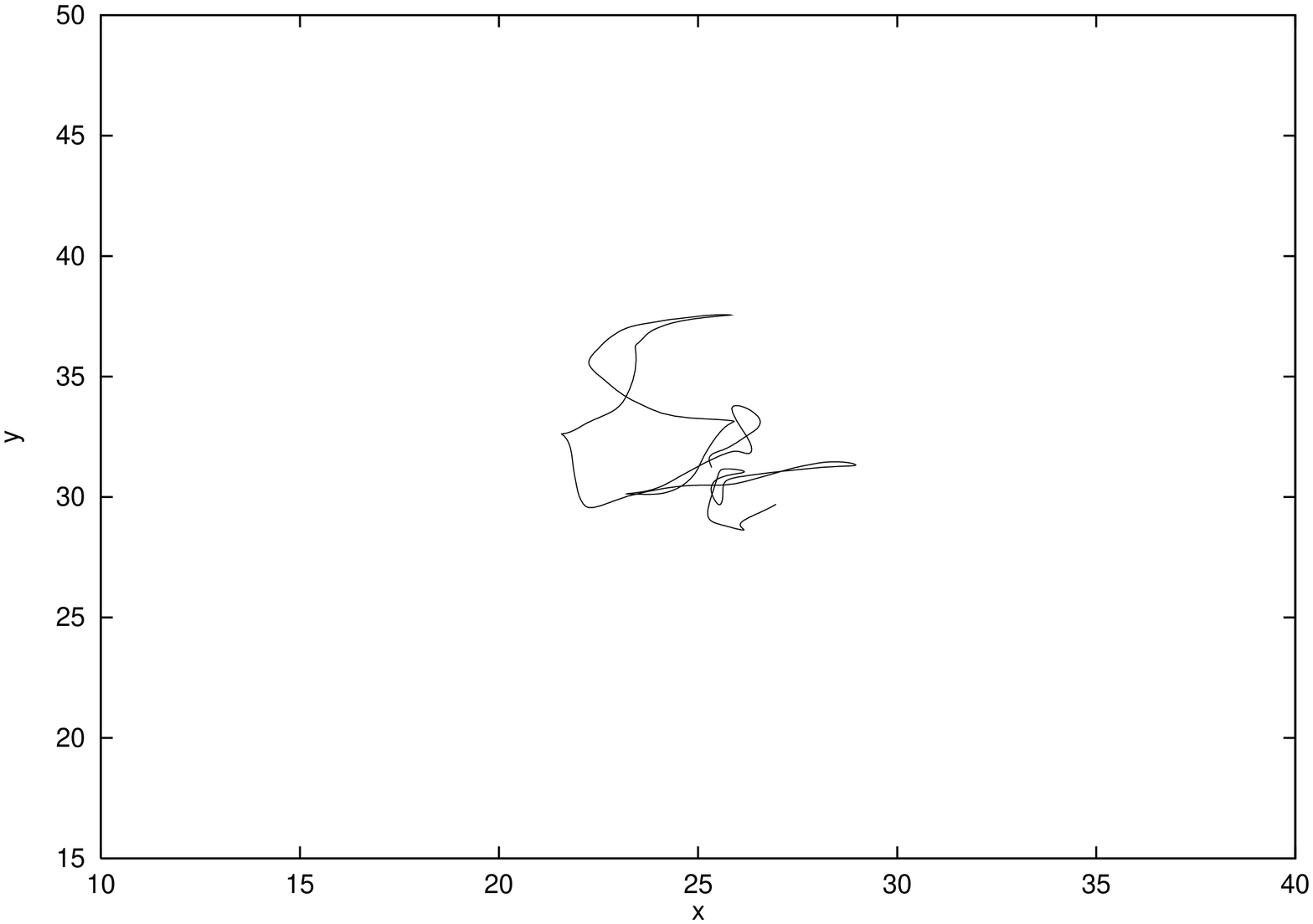,angle=270,height=6.5cm,width=6.5cm} 
\psfig{figure=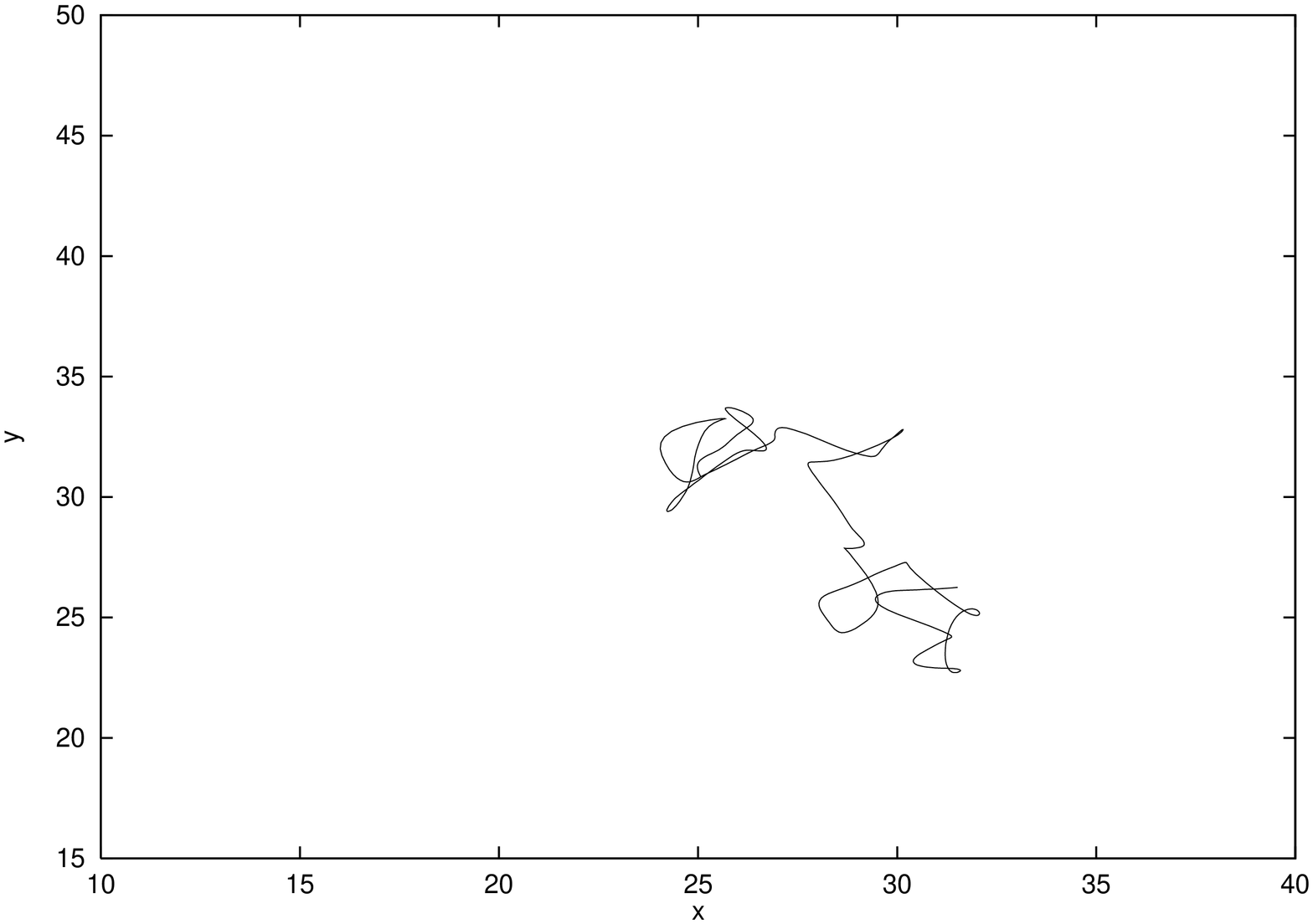,angle=270,height=6.5cm,width=6.5cm} 
}
}
\centerline{
\hbox{
\psfig{figure=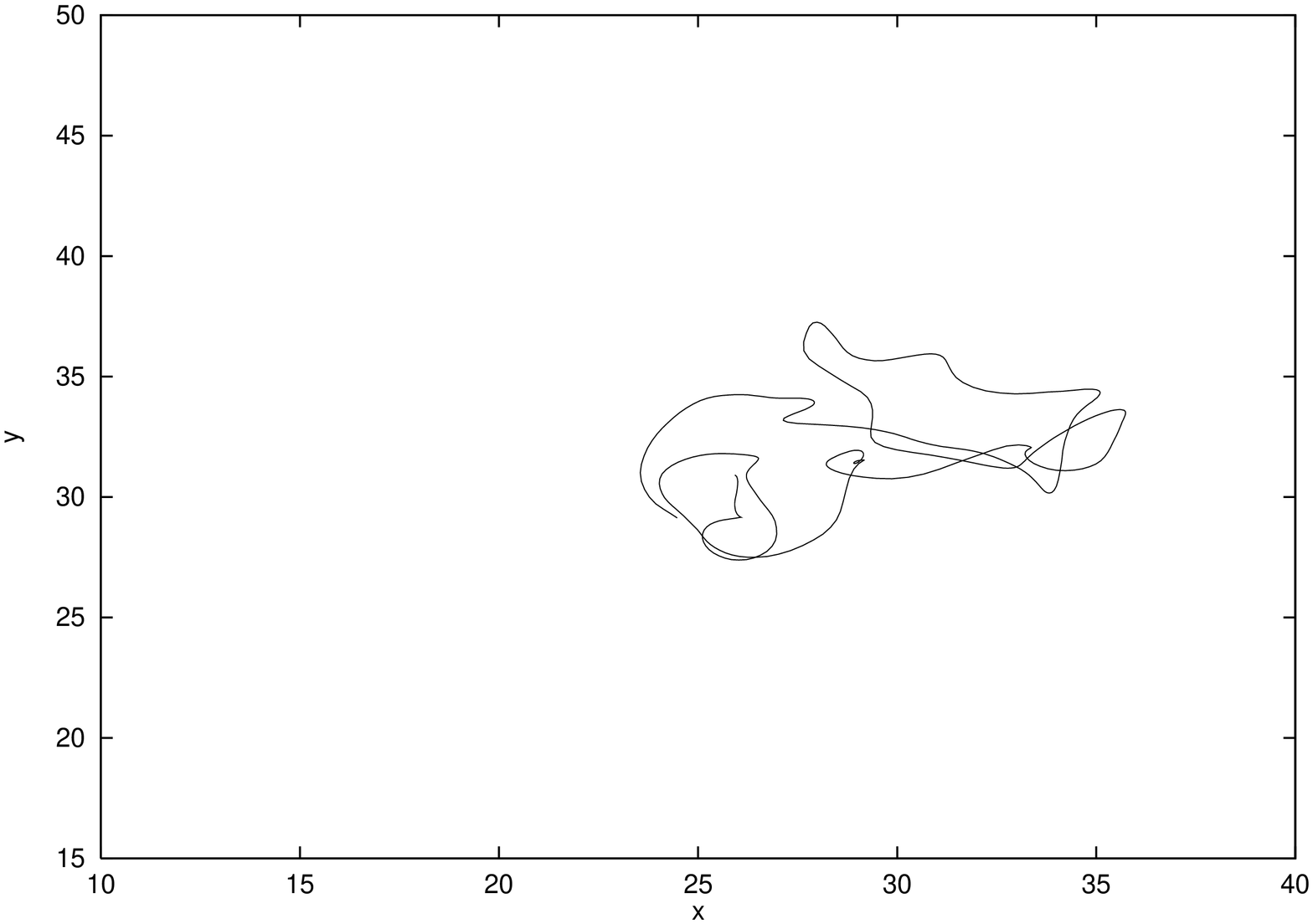,angle=270,height=6.5cm,width=6.5cm} 
\psfig{figure=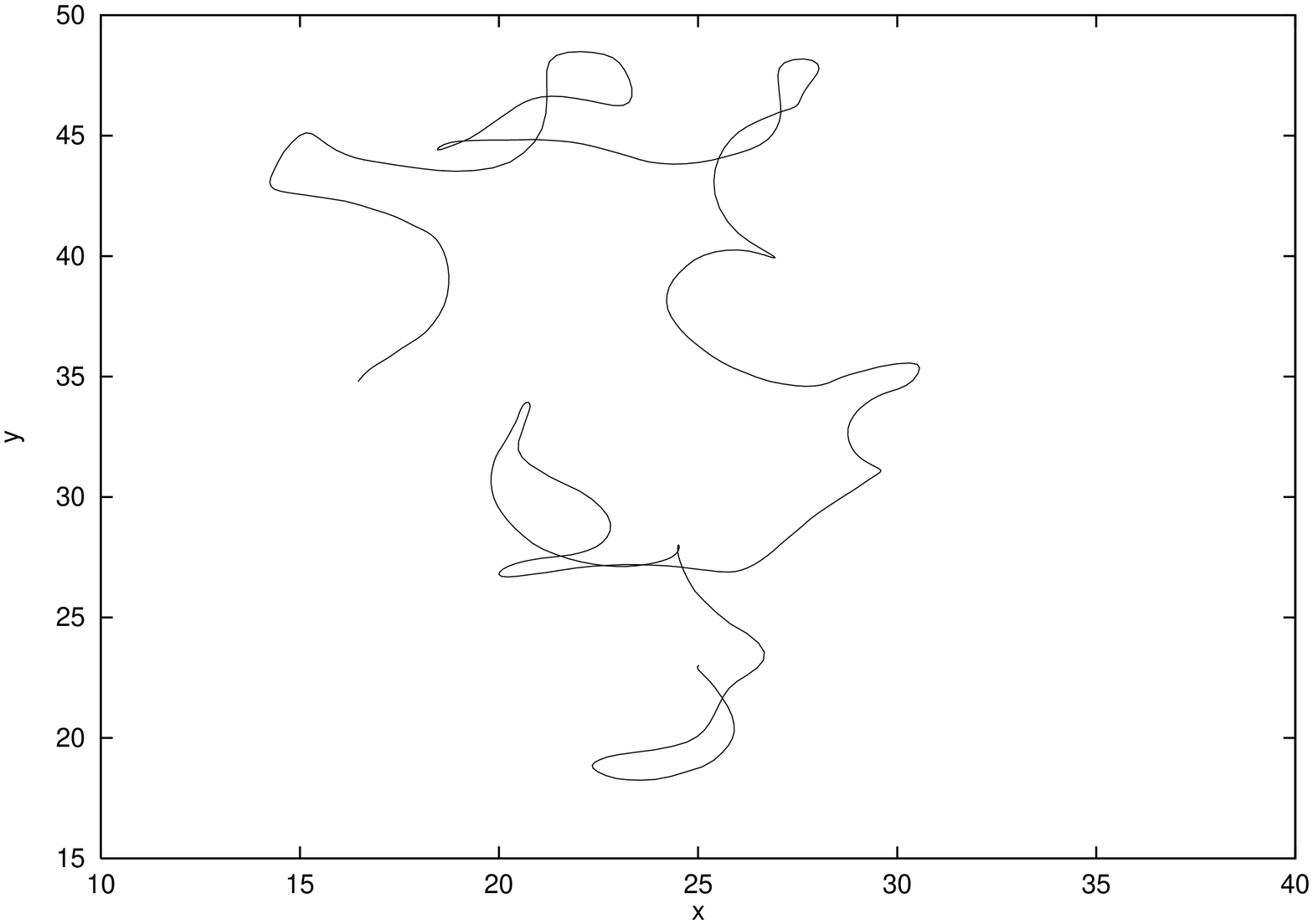,angle=270,height=6.5cm,width=6.5cm} 
}
}
\caption{
The trajectory of an advected particle for different values 
of \protect$\theta$ in the same $h$-field. 
From upper right to lower right 
\protect$\theta = \{0^{\circ},18^{\circ},72^{\circ},90^{\circ} \}$.
The particle moved for \protect$90$ time units. 
From the numerical solution of the 
2D Kuramoto-Sivashinsky equation, the field  $h$ is known only at discrete 
points in space.
 To find the values of the velocity field at all points in space, making
the $h$-field a continuous function we used bicubic 
spline interpolation \protect\cite{nr}.
 This is a standard method for interpolating a function 
which is only known at
discrete points. The interpolated function returned 
has continuous derivatives.
}
\label{fig:partraj}
\end{figure}

 In FIG.~\ref{fig:partraj} we show the trajectories of single particles 
 for different values of $\theta$ advected in the same $h$-field.
 
 It is seen how the trajectory corresponding to pure Hamiltonian motion 
 ($\theta=90^{\circ}$) is
made up of segments of smoothly rounded curves joined 
in sharp corners.
This is because in the changing field, the particle for some time spirals
around one maximum. When this maximum disappears, the particle changes
direction drastically and spirals
around another maximum.

The particle moving in a pure a gradient velocity field ($\theta=0^{\circ}$)
moves at a (local) maximum and thus follows the cell's motion.
When the cell is destroyed, it jumps to a new cell.

It is seen how the particles travel significantly longer in the Hamiltonian
($\theta=90^{\circ}$) case than in the gradient ($\theta=0^{\circ}$) case.

\subsection{Strange attractor}
For certain values of $\theta$
the particles actually move on a strange attractor.
 The range of the values of $\theta$ that gives
a strange attractor can
be estimated by calculating  the Lyapunov 
exponents for the particles.
Given  the two Lyapunov exponents of a dynamical system, of which at least
one should be positive, 
the Lyapunov dimension $D_L$ is given by \cite{ott}
\BE
D_L = 1 + \frac{\lambda_1}{|\lambda_2|}
\label{eq:dl}
\EE

The Lyapunov exponents are calculated in the following way 
\cite{benettin}:
We linearize the equations of motion of a
small disturbance $\delta {\bf x}(t)$ around the trajectory ${\bf x}(t)$.
The equations of motion for the system are given as:
\BE
\dot  x_i = V_i({\bf x},t)
\EE
We expand the equations of motion of a small disturbance 
$\delta{\bf x}(t)$ in a series around ${\bf x}(t)$ to first order and get
\BE
\dot{\delta{\bf x}_n(t)} = {\bf J({\bf x})} \cdot {\delta \bf x_n(t)}
\EE
where 
${\bf J({\bf x})} = \frac{\partial V({\bf x},t)}{\partial {\bf x}(t)}$
  is the Jacobian of the equations of motion
in the point ${\bf x}(t)$.
We use equations for two disturbances, 
$\delta{\bf x}_1(t)$ and $\delta{\bf x}_2(t)$,
since we want to find two Lyapunov exponents.
The initial conditions for the first disturbance vector are 
chosen in a random direction with length $|\delta {\bf x}_n(t)| = 1$.
The second disturbance vector is chosen orthonormal
to the first.
The equations of motion of the disturbances are integrated along 
with the equations of motion of the trajectory.
If the system is chaotic, at least one of 
the vectors $\delta {\bf x}_n(t)$ grow exponentially
and it is necessary to re-orthonormalize the vectors 
from time to time 
to avoid numerical overflow.
The re-orthonormalization is done in a way such that the first vector 
is simply renormalized while the other vector is turned and renormalized.
Let  $\tau$ be the time between 
consecutive re-orthonormalizations.
At each re-orthonormalization instant $j\tau$, the length $\alpha_j^{(1)}$ 
of the first vector  and the area $\alpha_j^{(2)}$ 
spanned by the two vectors are stored. 
The Lyapunov exponents are then given by 
\begin{eqnarray}
\lambda_1 &=& \lim_{t\rightarrow \infty} \frac{1}{l\tau}
\sum_{j=1}^{l} \ln \alpha_j^{(1)} \\ \nonumber
\lambda_2 &=& - \lambda_1 + \lim_{t\rightarrow \infty} \frac{1}{l\tau}
\sum_{j=1}^{l} \ln \alpha_j^{(2)} 
\end{eqnarray}

The Kaplan-Yorke conjecture states that $D_L$ actually gives the information 
dimension $D_1$ of the attractor of the system.

We see that for the chaotic Hamiltonian case which has
one positive and one negative Lyapunov exponent with the same numerical value,
 $D_L=2$, and thus the attractor of the particles is a two dimensional
 subset of the whole plane, as expected for a Hamiltonian system.
It is thus necessary to introduce some dissipation (i.e.
 $\theta<\ 90^{\circ}$) to find a strange attractor.

\begin{figure}[t]
\centerline{
\psfig{figure=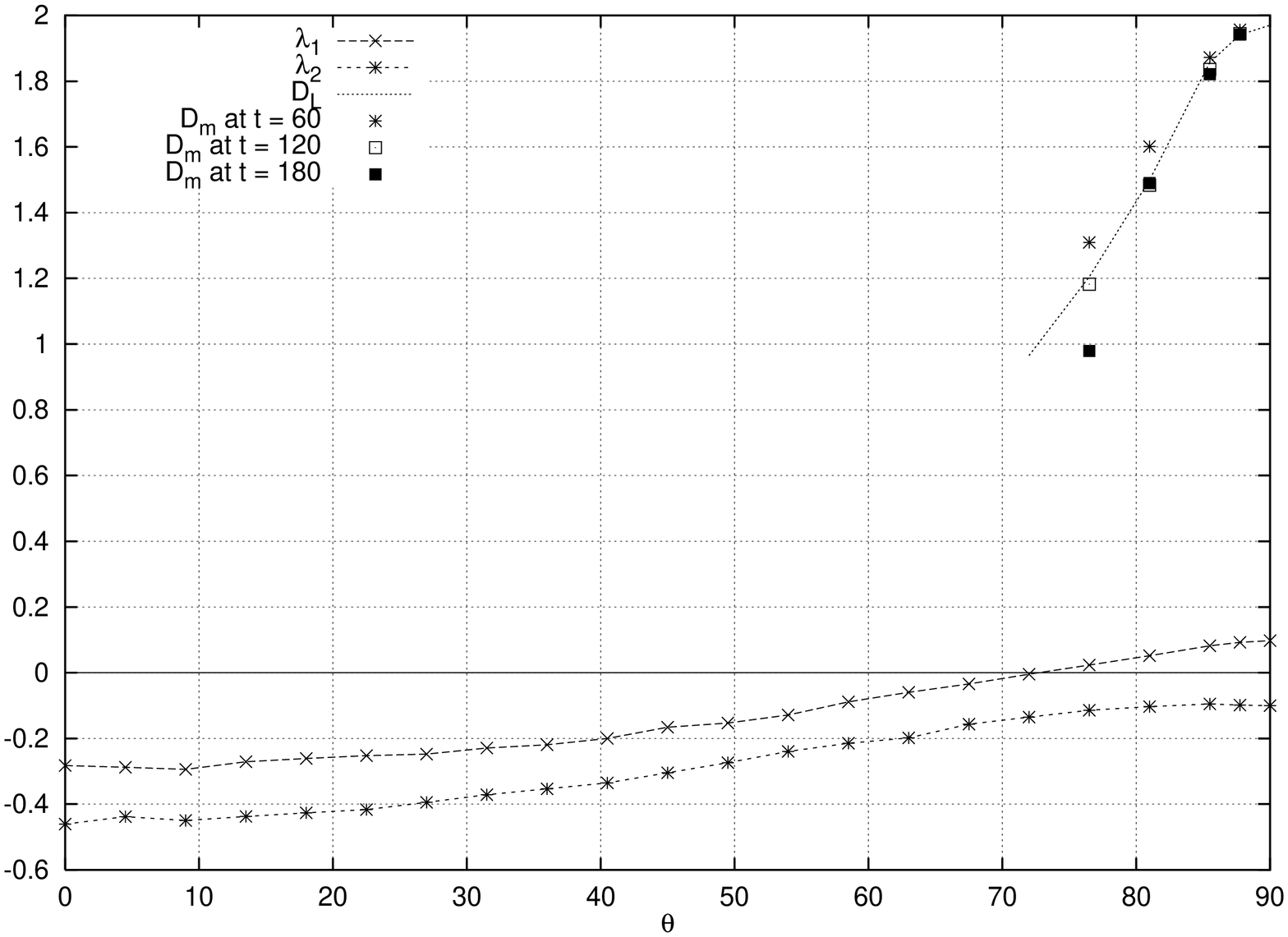,angle=270,height=8cm,width=13.5cm}
}
\caption{
The two Lyapunov exponents of the particle motion 
for different values of $\theta$ are shown as the two lower curves.
The upper curve show the Lyapunov dimension as calculated from the 
Kaplan-Yorke conjecture when one Lyapunov exponent is positive.  
The data points are the measured dimensions of the
attractors from the simulations}
\label{fi:2dlyap}
\end{figure}

The information dimension is defined as
\BE
D_1 = - \lim_{\epsilon \rightarrow 0} 
\frac{\sum_i^N p_i\log p_i}{\log \epsilon},
\label{eq:d1}
\EE
where $p_i$ is the natural measure of particles in box $i$.
The natural measure in a box is the fraction of the
total number of particles, or normalized density, in each box.
The sum is taken over all boxes $i$, which has nonzero natural measure.
In FIG.~\ref{fi:2dlyap} numerically calculated values of the two 
 Lyapunov exponents are shown as a function of the angle $\theta$ (the two 
 lower curves).
 It is seen that $\theta$ has to be quite large, $\theta \approx 72^{\circ}$, 
 to obtain one positive
 Lyapunov exponent.
 Also shown is the Lyapunov dimension of the corresponding attractor.

 It should be noted, that the Lyapunov exponents and their signs are
also strongly dependent on the ratio of the time scales, i.e., the 
values of the constant $a$. 
The Lyapunov exponents decrease with increasing $a$ and the
$\theta$ at which the transition 
 $\lambda_1 > 0$ occur increases.
In the limit of $a \rightarrow \infty$, the transition 
newer occurs although 
both Lyapunov exponents in the pure Hamiltonian system 
$(\theta = 90^{\circ})$
are zero corresponding to a $h$-field that 
is constant in time i.e. $h=h(x,y)$.
\subsection{Results for the KS Equation}
In FIG.~\ref{fi:sommerer} we show  
pictures of the strange attractor
on the surface of a real fluid, taken from
\cite{sommererscience}.
It is seen that there are not that many large scale structures, 
the large bends of
the strange attractor are big compared to the image size.
We thus expect the system size of the Kuramoto-Sivashinsky equation
needed to produce a strange attractor to be only a few cells big.
It of course has to big enough to be chaotic.
A system of size $32 \times 32$ spatial units were chosen.
This corresponds to approximately 4 cells in each direction.
In the experiments $\approx 50$ million particles were used 
\cite{sommererscience}. 
We thus have
to advect many particles to see the strange attractor.
As a compromise between computing time and the large number needed, 
$190 000$ particles were used. 
The particles were originally placed in a square grid covering the whole
system. The 2D-Kuramoto-Sivashinsky equation was integrated 
using an explicit finite difference scheme with grid
spacings of half a spatial unit, thus $64 \times 64$ mesh point were used. 
The gradient field between the mesh points was found using a bicubic spline
interpolation \cite{nr}.
The computing time for the integration of the system with particles 
was one hour per time unit on the used computers (Pentium 166
MHz.)

\begin{figure}[t]
\centerline{
\hbox{
\psfig{figure=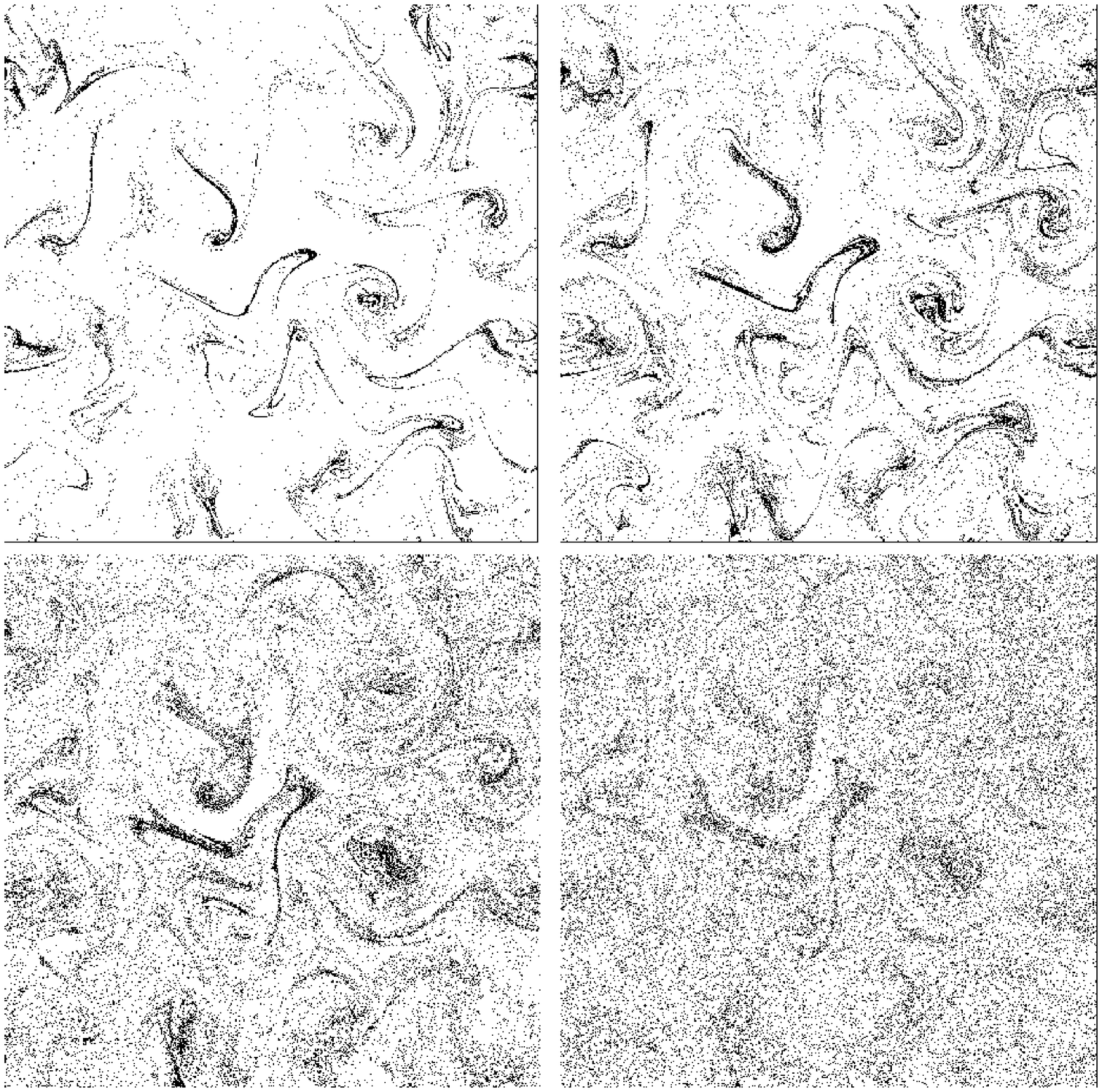,height=13cm,width=13cm} 
}
}
\caption{
A realization of the particle patterns at time $t=60$.
Values of $\theta$ from upper left to lower right: 
$76.5^{\circ}, 81.0^{\circ}, 85.5^{\circ}, 87.75^{\circ}$.
The measured dimensions are from upper left to lower right 
$ 1.309 \pm 0.005, 1.601 \pm 0.008, 1.873 \pm 0.006, 1.957 \pm 0.005$.
}
\label{fi:2dpate}
\end{figure}

\begin{figure}[t]
\centerline{
\hbox{
\psfig{figure=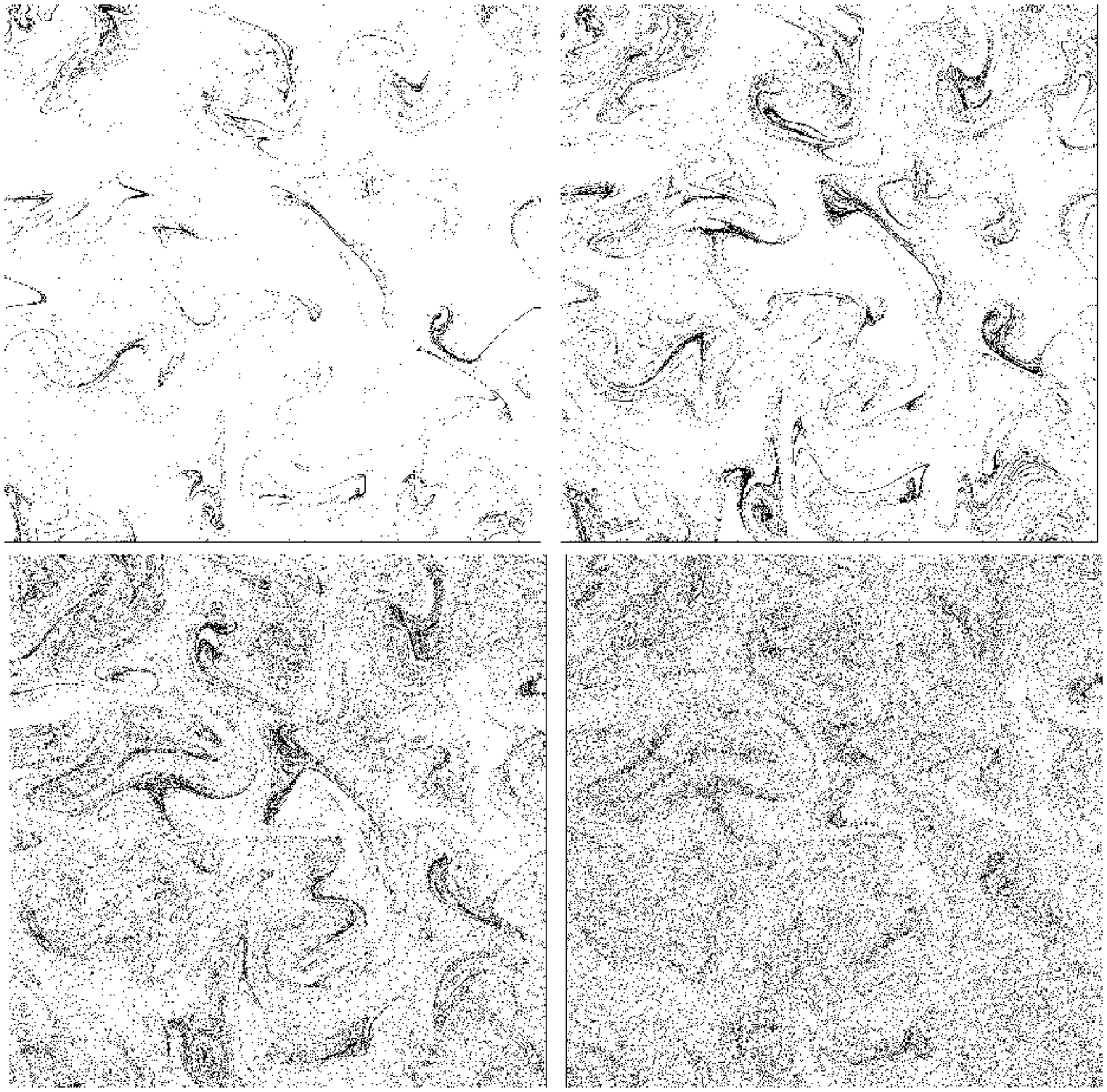,height=13cm,width=13cm} 
}}

\caption{
A realization of the particle patterns at time $t=180$.
Values of $\theta$ from upper left to lower right: 
$76.5^{\circ}, 81.0^{\circ}, 85.5^{\circ}, 87.75^{\circ}$.
The measured dimensions are from upper left to lower right 
$ 0.979 \pm 0.007, 1.49 \pm 0.008, 1.821 \pm 0.004, 1.942 \pm 0.006$.
}
\label{fi:2dpatg}
\end{figure}

After a transient time the particles produce a pattern 
which resembles a strange attractor
which changes in time. Patterns for different values of $\theta$
but the same $h$-field 
at the same time are shown in FIG.~\ref{fi:2dpate}-\ref{fi:2dpatg}.

We have estimated the information dimensions of the patterns numerically, using
the information dimension $D_1$ (\ref{eq:d1}), 
where the $32 \times 32$ field is divided into $N$ square boxes 
with side-length $\epsilon$. 
The dimensions are found as the slope of a straight line fit to a plot 
of $\sum_i^N p_i\log p_i$ versus $\log \epsilon$.

The measured information dimensions with error bars 
from the fit,
the Lyapunov exponents and the  
Lyapunov dimensions calculated from 
(\ref{eq:dl}) are shown in FIG.~ \ref{fi:2dlyap} as functions of
$\theta$.

The Lyapunov exponents change continuously with $\theta$
and with that the information dimension and we can thus 
 choose the dimension of the attractor by varying $\theta$.

\begin{figure}[t]
\centerline{
\mbox{
\psfig{figure=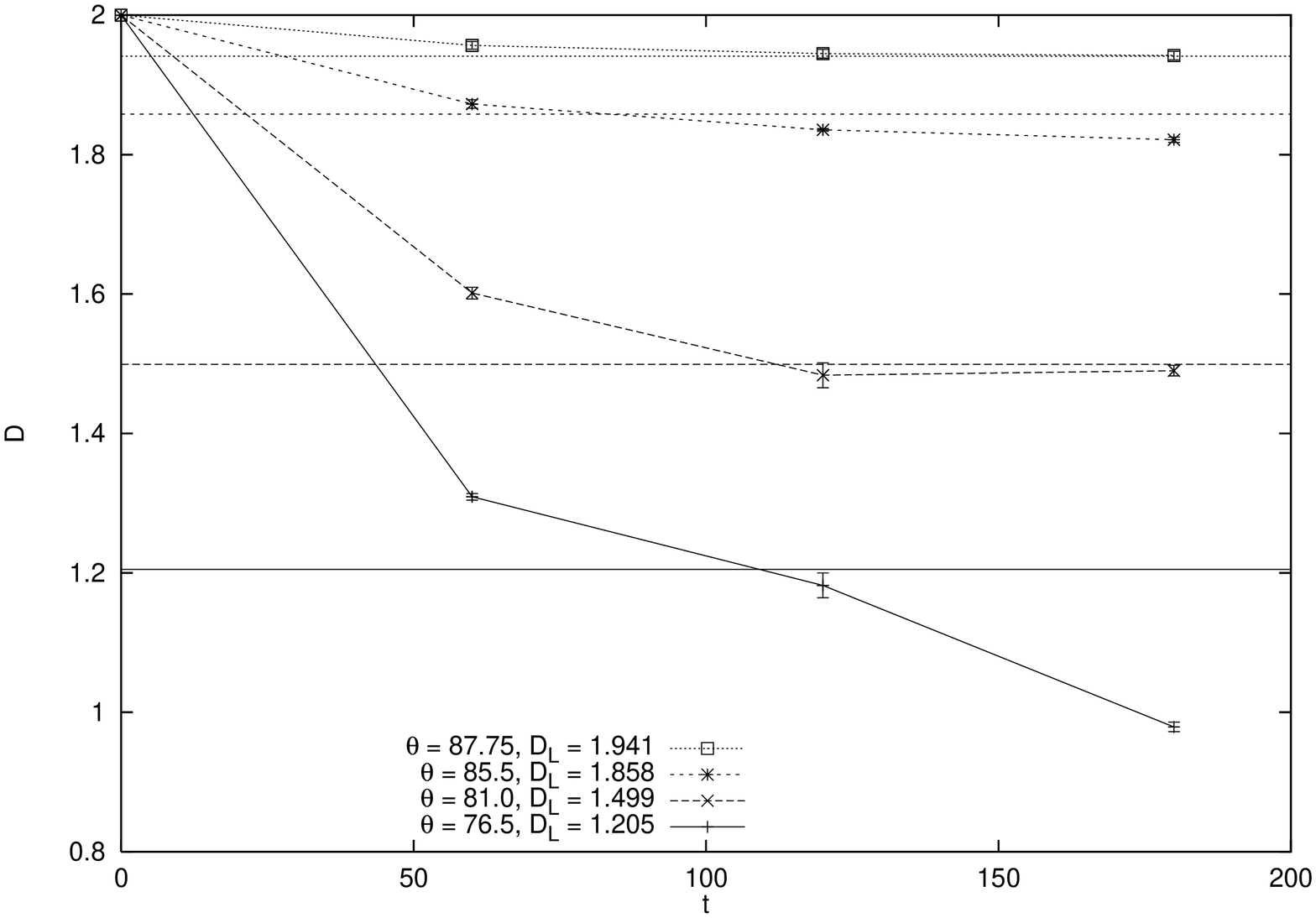,angle=270,height=8cm,width=13cm}
}}
\caption{
The measured information dimension at different times 
for each of the four used values of $\theta$.
The horizontal lines are the corresponding calculated Lyapunov dimensions
$D_L$.
It is seen how the values of the measured dimensions 
apparently converge to the calculated Lyapunov dimensions 
for the three highest values of $\theta$.
For the lowest value of $\theta$, near the threshold of chaos,
the dimension of the particle distribution 
 apparently does not converge in time.
The fits could seldomly be done over more than one
decade, due to cross-over effects at each end of the interval. The limitations
are mainly due to computional problems of handling more than the 
190.000 particles used.
}
\label{fi:convdim}
\end{figure}

In FIG.~\ref{fi:convdim} we show how the measured value of the information
dimension changes in time. 
The horizontal lines in the plot are the corresponding 
calculated Lyapunov dimensions $D_L$.
For the three highest used values of  
$\theta$, the information dimensions are seen 
to converge nicely to the calculated Lyapunov dimensions,
while for the lowest used value of $\theta$, 
the information dimension is decreasing in time.
This is probably a numerical problem connected
with the existence of local regions where the velocity field is 
contracting in all directions for a long time, which causes
particles to coalesce and thus lowers the information dimension.

\section{Conclusion}

We have found fractal tracer distributions of particles advected in the
velocity field of the 2D-Kuramoto-Sivashinsky equation.
The measured dimensions of attractors are in good agreement 
with calculated Lyapunov dimensions.
It could be interesting to see if the fractal tracer distributions could be
obtained for particles advected in other equations exhibiting
spatio-temporal chaos, e.g. the
Complex Ginzburg-Landau equation, where some work on 
the motion of advected particles has been
done \cite{elsebeth}.

\section{Acknowledgment}
The authors thank 
G. Huber, E. Ott and A. Pikovsky
for discussions.

\newpage

\newpage 
\newpage

\newpage
\newpage

\newpage


\begin{thebibliography}{10}

\bibitem{sommererscience}
J.~C. Sommerer and E.~Ott.
\newblock Particles floating on a moving fluid: A dynamically comprehensible
  physical fractal.
\newblock {\em Science}, 259:335--339, 1993.

\bibitem{sommerer}
J.~C. Sommerer.
\newblock Fractal tracer distributions in complicated surface flows: an
  application of random maps to fluid dynamics.
\newblock {\em Physica D}, 76:85--98, 1994.

\bibitem{LOS1}
L.~Yu, E.~Ott, and Q.~Chen.
\newblock {\em Phys. Rev. Lett.}, 65:2935, 1990.

\bibitem{LOS2}
L.~Yu, E.~Ott, and Q.~Chen.
\newblock {\em Physica D.}, 53:102, 1991.

\bibitem{kuramoto}
Y.~Kuramoto.
\newblock {\em Chemical Oscillations, Waves, and Turbulence}.
\newblock Springer Verlag, 1984.

\bibitem{sivashinsky}
G.~I. Sivashinsky.
\newblock Nonlinear analysis of hydrodynamic instability in laminar flames--i.
  derivation of basic equations.
\newblock {\em Acta Astronautica}, 4:1177--1206, 1977.

\bibitem{ksdif}
T.~Bohr and A.~Pikovsky.
\newblock Anomalous diffusion in the {K}uramoto-{S}ivashinsky equation.
\newblock {\em Phys. Rev. Lett.}, 70:2892, 1993.

\bibitem{ott}
E.~Ott.
\newblock {\em Chaos in dynamical systems}.
\newblock Cambridge University Press, 1993.

\bibitem{benettin}
G.~Benettin, L.~Galgani, A.~Giorgilli, and J.-M. Strelcyn.
\newblock {L}yapunov {C}haracteristic {E}xponents for {S}mooth {D}ynamical
  {S}ystems and for {H}amiltonian {S}ystems: A {M}ethod for {C}omputing {A}ll
  of {T}hem. {P}art 2: {N}umerical {A}pplication.
\newblock {\em Meccanica}, 15:22, 1980.

\bibitem{nr}
W.H. Press, S.A. Flannery, B.P.and~Teukolsky, and W.T. Vetterling.
\newblock {\em Numerical Recipes in C}.
\newblock Cambridge University Press, second edition, 1992.

\bibitem{elsebeth}
G.~Huber, E.~Schr{\"o}der, and P.~Alstr{\o}m.
\newblock Self-diffusion and relative diffusion in defect turbulence.
\newblock {\em Physica D}, 96:1--8, 1996.


\end{thebibliography}
\end{document}